\documentclass[prd, reprint, amsmath, amssymb, superscriptaddress, nofootinbib, showpacs, aps, floatfix]{revtex4-1}

\usepackage[justification=justified,font=small,labelfont=bf]{caption}

\newcommand{\macro}[1]{\texttt{\textbackslash#1}}
\newcommand{\m}[1]{\macro{#1}}

\newcommand{\Chi}{\mathrm{X}}
\newcommand{\Kappa}{\mathrm{K}}
\usepackage{graphicx}%
\usepackage{multirow}
\usepackage{float}
\usepackage{epstopdf}
\usepackage[utf8]{inputenc}
\usepackage[english]{babel}
\usepackage{footnote}
\usepackage{xcolor}
\usepackage{textcomp}
\usepackage{color} 
\usepackage{soul}
\usepackage{siunitx}
\usepackage[capitalise,noabbrev]{cleveref}

\usepackage[skip=6pt plus1pt, indent=12pt]{parskip}
\usepackage{silence}
\begin{document}

\title{Cosmogenic Muon Background Characterization for the Colorado Underground Research Institute (CURIE)}%

\author{Dakota K. Keblbeck}\thanks{Corresponding author}
\affiliation{Department of Physics, Colorado School of Mines, Golden, Colorado 80401, USA}

\author{Eric Mayotte}
\affiliation{Department of Physics, Colorado School of Mines, Golden, Colorado 80401, USA}

\author{Uwe Greife}
\affiliation{Department of Physics, Colorado School of Mines, Golden, Colorado 80401, USA}

\author{Kyle G. Leach}
\affiliation{Department of Physics, Colorado School of Mines, Golden, Colorado 80401, USA}

\author{Wouter Van De Pontseele}
\affiliation{Department of Physics, Colorado School of Mines, Golden, Colorado 80401, USA}
\affiliation{Laboratory for Nuclear Science, Massachusetts Institute of Technology, Cambridge, Massachusetts 02139, USA}

\date{\today}%

\begin{abstract}
We present the characterization of cosmogenic muon backgrounds for the Colorado Underground Research Institute (CURIE), located in the Edgar Experimental Mine (EEM) in Idaho Springs, Colorado. The CURIE facility at the EEM offers a versatile shallow underground environment, with accessible horizontal tunnel access and stable rock formations ideal for low-background physics experiments. We have measured the total underground muon flux in two locations, Site 0 and Site 1, yielding values of $\phi$ = 0.246 $\pm$ 0.020$_{sys.}$ $\pm$ 0.012$_{stat.}$ and 0.239 $\pm$ 0.025$_{sys.}$ $\pm$ 0.010$_{stat.}$ $\mu\text{/}$\unit[per-mode = symbol]{\square\metre\per\second}, respectively. We have utilized GEANT4 and PROPOSAL Monte Carlo simulations with Daemonflux and MUTE to model the muon flux at both sites, as well as an additional future location. We find good agreement between measurement and simulations, demonstrating the first instance of this computational framework being successfully used for depths $<$ 1 km.w.e. The measured underground flux corresponds to a factor of 700 reduction compared to the sea level flux. Additionally, we present a new depth-intensity relationship to normalize the mountain overburden to an equivalent flat depth, enabling direct comparison with other underground facilities. We report an average equivalent vertical depth of 0.415 $\pm$ 0.027 km.w.e. Based on our measurements, this work highlights the facility's capability for hosting low-background experiments, addressing the demand for shallow underground research spaces.
\end{abstract}

\maketitle

\section{Introduction}
When high-energy primary cosmic rays interact with Earth's atmosphere, showers of secondary particles are produced, including cosmic ray muons. These muons are primarily the product of $\pi$ and $\Kappa$ meson decay, with energies spanning several orders of magnitude~\cite{mu_atmo_energies}. At sea level, muons are the most abundant charged particles, with an average flux of  $\approx 1 \mu\text{/}$\unit[per-mode = symbol]{\per\centi\square\metre\per\min} and a mean energy of $\approx$ \SI{4}{\GeV} \cite{mu_flux, mu_energy}. For physics experiments on the surface, these muons are a common source of background noise, collectively referred to as muon-induced backgrounds. These backgrounds are due to the direct interaction of muons with electronics, the production of secondary radiation from muon interactions in surrounding materials (e.g., neutron production from spallation), and the interactions of the products of muon decay.

To mitigate the effect of cosmic ray muons as a background source, many fundamental physics experiments that aim to detect small signals or pursue rare event searches rely on low-background underground research facilities \cite{dune_mu, nexo_mu, cdms_mu}. The overhead rock, or overburden, at these facilities provides the shielding to attenuate the muons. Although relatively rare, these facilities exist worldwide and are typically classified based on their muon-flux-normalized depth in kilometer-water-equivalent (km.w.e.) shielding as either ``shallow-underground" (depths $<$ 1\,km.w.e.) or ``deep-underground" (depths $>$ 1\,km.w.e.). The overburden of these underground facilities is one of two types: flat or mountain. For labs with a flat overburden, the muon flux is assumed to be symmetric with respect to the azimuthal angle. In contrast, labs underneath mountains have a complex overburden and, therefore, a muon flux that depends on both zenith and azimuth angles. To directly compare these two types of underground facilities, empirical models \cite{mei_hime,mitrica, macro, sudbury, lvd} have been developed to convert the km.w.e.\ depth for mountain sites to an equivalent depth relative to a flat overburden.

There are approximately a dozen deep underground labs in the world; of these, five are in North America: Soudan (USA)~\cite{soudan}, WIPP (USA)~\cite{wipp}, KURF (USA)~\cite{kurf}, Sanford (SURF) (USA)~\cite{surf}, and SNOLAB (Canada)~\cite{snolab}. The high demand from large-scale experiments, such as those in Refs.~\cite{dune_mu, nexo_mu, cdms_mu}, often makes staging small-scale experiments in these facilities difficult. Therefore, effective shallow underground facilities can be more cost-effective and practical for many low-background and small- to mid-scale projects, as they may be able to tolerate moderate muon fluxes while benefiting from the complete rejection of other types of cosmogenic secondaries.


Over the last few decades, shallow facilities have been used effectively and productively in the United States (U.S.) and Europe. However, few such facilities exist, and this constraint can limit the development of critical technologies that require these spaces. Currently, in the U.S., two major facilities used for basic and applied shallow underground science are located at Department of Energy (DOE) National Laboratories: the Pacific Northwest National Laboratory (PNNL) in Richland, WA, and Fermilab in Batavia, IL. The PNNL facility is at a depth of \SI{30} m.w.e.\ and provides a muon reduction of $\approx 6\times$  compared to sea level~\cite{pnnl}. Depending on the location, the Fermilab facility offers an overburden of $\approx$ \SI{225} - \SI{300} m.w.e.\ and muon reductions of $\sim 200\times$ to $400\times$ compared to sea level~\cite{Figueroa2024, temples}. These two U.S-based facilities provide natural protection from cosmic muons where sensitive experiments and novel technology development can be hosted. Given the expanding research areas requiring these capabilities, this provides the motivation to develop new shallow facilities that can provide more space, operational flexibility, similar or better muon shielding, and easy site access.

Here, we present the analysis and characterization of the cosmogenic muon background for the new Colorado Underground Research Institute (CURIE) under development in the Edgar Experimental Mine (EEM)~\cite{edgarmine}, at the Colorado School of Mines.


\section{Edgar Experimental Mine}\label{sec:EEM}
\begin{figure}
    \centering
    \includegraphics[width = 8.5cm]{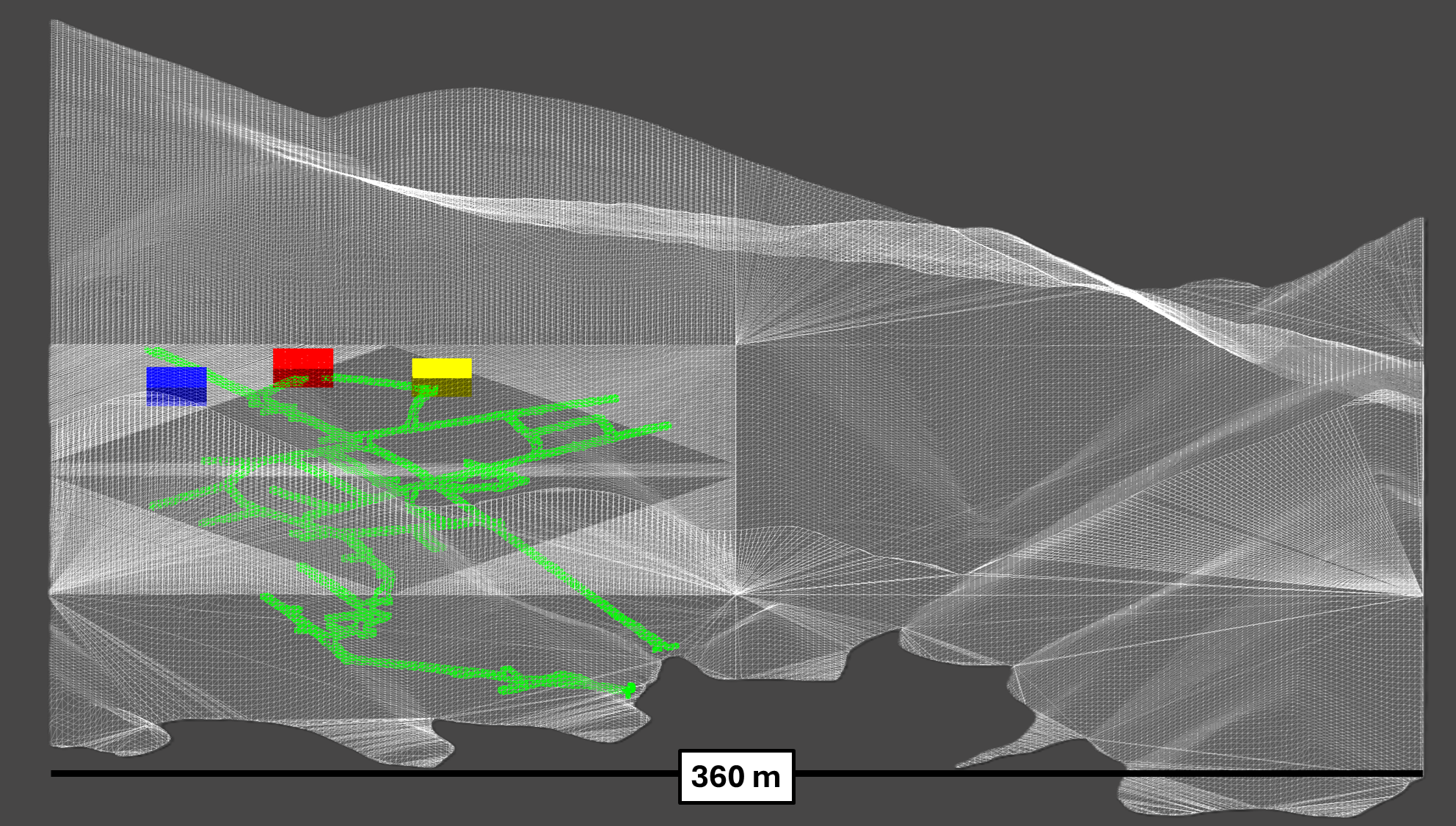}
    \caption{3D rendering of the EEM topology with the mine access ways shown in green. The relative positions of Site 0 (red), Site 1 (blue), and Site 2 (yellow) are also shown.}
    \label{fig:eem_miami_tunnel}
\end{figure}

The EEM is located in Idaho Springs, CO, at an elevation of \SI{2401}{\m} above sea level. It has two main operational areas that connect internally: the Army tunnel and the Miami tunnel. The Miami tunnel is the direct, shortest access to the research sites discussed in this work. \cref{fig:eem_miami_tunnel} shows the extent of the access tunnels, which provide horizontal access to the lab sites. The EEM was originally a silver and gold mine in the 1870s, after which it was acquired by the Colorado School of Mines for use as a research and training facility for mining engineering students. Since then, it has seen substantial expansions and improvements in the underground working facilities, most notably: \SI{110}{\V} single- and \SI{440}{\V} three-phase power supplies, an extensive \SI{1275}{\m^3/min} ventilation system, and compressed air and water sources. Additionally, the EEM is excavated from Gneiss rock, which is analogous to granite but has foliation, making it a very stable and safe mine location from a rock dynamics standpoint~\cite{gneiss_rock}.


The stability of the rock formation allows for a high level of flexibility in choosing underground locations for potential research sites. Three lab sites are used throughout the analysis in this paper: the preexisting site Bureau of Mines (BOM) Stope (Site 0), the newly excavated Site 1 (completed), and Site 2 (in preparation). These sites are located off the Miami tunnel, and their relative positions can be seen in \cref{fig:eem_miami_tunnel}. They are located $\sim 400$ horizontal meters into the Miami tunnel, with tunnel openings varying from around $\SI{1.8}{\m} \times \SI{1.8}{\m}$ at the narrowest to $\SI{4.5}{\m} \times \SI{4.5}{\m}$ at the widest. The narrowest access through the Army tunnel is approximately $\SI{2.4}{\m} \times \SI{2.4}{\m}$. The vertical height to the surface is $\approx$ \SI{200}{m} for all sites. The Miami tunnel is approximately level over the \SI{400}{\m} to the sites and features a rail line with electric and diesel locomotives. Site 2 is tailored to house a cryogenic low-background facility.  
The facility will focus on developing high-resolution, low-threshold detectors for fundamental science. Additionally, the low background environment will allow for the controlled study of the effects of ionizing radiation, electromagnetic disturbances, and vibrations on superconducting circuits~\cite{PhysRevB.110.024519}, including qubits and quantum-adjacent technologies such as cryo-CMOS circuits~\cite{cryocmos}.

\section{Muon Flux Simulations}\label{sec:SimAnalysis}
Several Monte Carlo toolkits exist to simulate the cosmogenic muon background at underground research facilities~\cite{g4, fluka, music}. In this work, we used Daemonflux (DAta-drivEn MuOn-calibrated atmospheric Neutrino Flux) v0.8.1 \cite{daemonflux, daemonflux2} to simulate the muon flux at the surface, and MUTE (MUon inTensity codE) v2.0.1 \cite{mute, mute2} which takes muons from the surface flux model and propagates them to the simulated underground sites.

Daemonflux is a Matrix Cascade Equation (MCEq) project which computes inclusive atmospheric lepton fluxes by using one-dimensional coupled cascade equations from Sibyll 2.3c~\cite{sibyll2.3c} for the surface muon fluxes. Detailed information on MCEq can be found in Refs. \cite{mceq2, mceq3}. Daemonflux is calibrated using the atmospheric flux Global Spline Fit (GSF) \cite{gsf} model and the inclusive hadronic interaction Data-Driven Model (DDM) \cite{mceq2}. The GSF model combines direct space and high-altitude balloon measurements, while DDM is a set of inclusive particle production cross-section measurements from fixed target accelerators. MUTE takes the surface fluxes from Daemonflux and convolves them with PROPOSAL (Propagator with optimal precision and optimized speed for all leptons) \cite{proposal, proposal2} to simulate the propagation of muons through the rock overburden. An accurate model of the rock overburden is required as input to simulate the muon flux.

\subsection{Modeling the overburden at the EEM}\label{sec:mtnprofiles}
\begin{figure}
    \centering
    \includegraphics[width = 8.5cm]{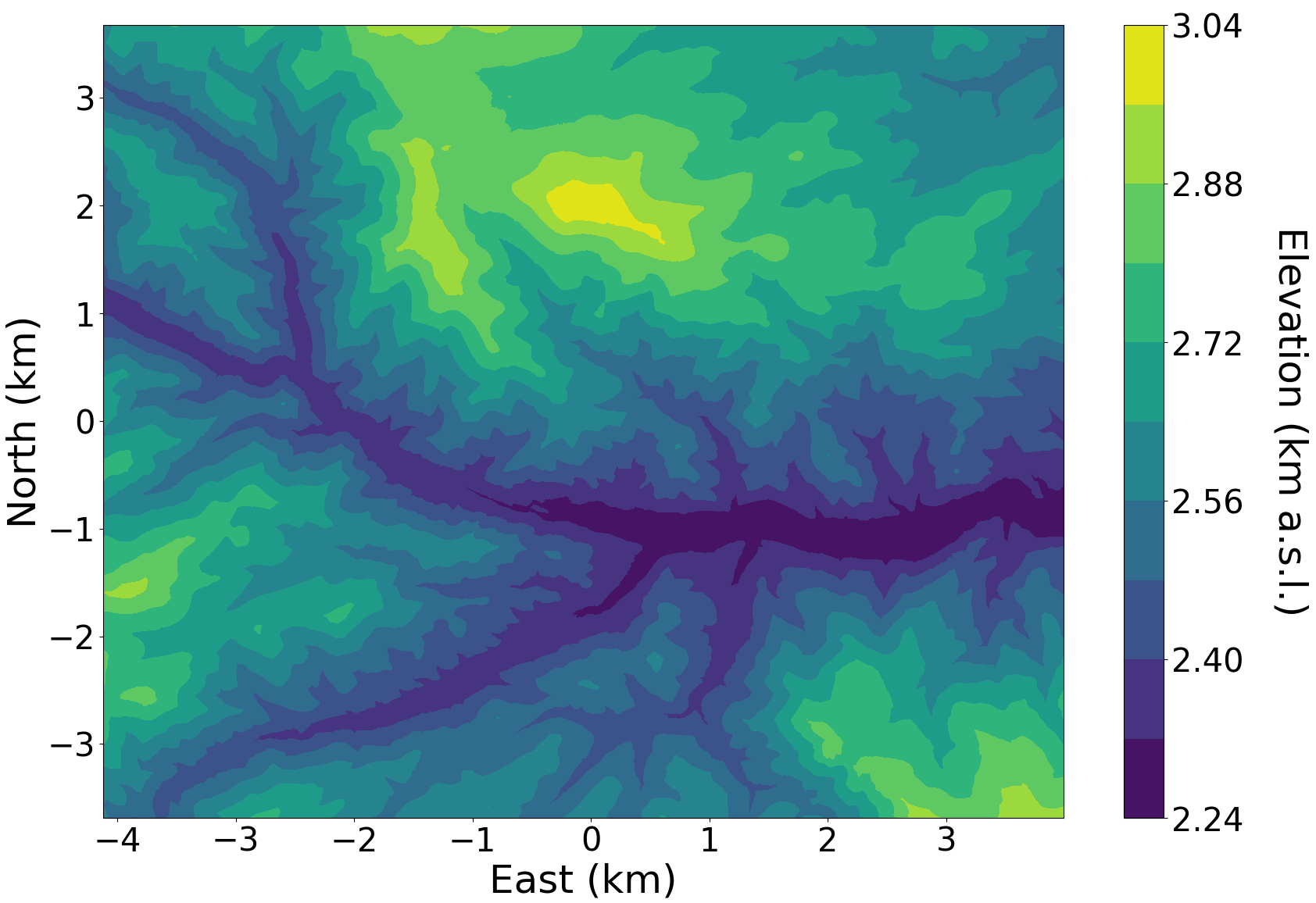}
    \caption{Topological map of the \SI{84}{\km^{2}} area extracted from the USGS DEM data. The map is centered such that the entrance to the EEM is located at the origin of the plot. The z-axis is given in km above sea level.}
    \label{fig:topo_map}
\end{figure}

To model the rock overburden above the low background facilities, the United States Geological Survey (USGS) Digital Elevation Model (DEM) data \cite{USGS2020} was acquired for an area of $\approx$ \SI{84}{\km^2} around the EEM. The USGS data at this site has an accuracy in the $x-y$ plane of \SI{1}{\m} \cite{usgs_xy} and a vertical accuracy of \SI{13.6}{\cm} \cite{USGS2020}. The DEM data was then extracted as $x$, $y$, and $z$ coordinates, corresponding to East, North, and altitude (Long: -105.52 E, Lat: 39.74 N, Alt: \SI{2401.82}{\m}), using the open-source software Quantum Geographic Information System (QGIS)~\cite{QGIS}. Figure \ref{fig:topo_map} shows a topological map of the surface elevation for the \SI{84}{\km^2} area centered on the Edgar Experimental Mine.

Site-centered surface models were created for each site. The Cartesian surface maps were converted to spherical coordinates $(\theta, \phi, R)$, where $R$ is the radial distance from the lab to the surface for a given zenith, $\theta$, and azimuth, $\phi$, angle combination.
For the muon propagation simulations, the $R$ values were converted to slant depth ($\Chi$) in units of km.w.e, via $\Chi = R*(\rho_{r}/\rho_{w}$), where $\rho_{r}$ is the density of the rock and $\rho_{w}$ is the density of water (\SI{0.997}{\g/\cm^{3}} at \ang{25} C). The final overburden model, from now on called the mountain profile, is stated in the coordinates ($\theta, \phi, \Chi$). Each mountain profile has more than 3 million data points, with randomly distributed spatial errors. Given this, the contribution of the statistical uncertainty from the USGS data is negligible compared to the geological systematic errors outlined below. 

\subsubsection{Rock Composition}\label{sec:rock}
The non-uniformity of rock can represent a large source of systematic uncertainty in the simulation of underground muon fluxes~\cite{rock_comp, rock_comp2}. This is because they modify the zenith and azimuth-dependent density profile of the overburden. These non-uniformities include the inhomogeneous distributions of rock layers' of different chemical compositions, the degree of water saturation in the rock, and the possible presence of large water deposits or voids. As such features are hidden behind meters of solid rock, quantifying the systematic uncertainties from the rock composition is difficult. 

To estimate these uncertainties, the USGS and Colorado rock composition and geologic survey data \cite{usgs_rock, colo_rock} for an area of $\approx$ \SI{20}{\km^{2}} surrounding the EEM was used to create 3 density profiles for each site yielding 9 total mountain profiles. The density profiles were a standard rock scenario ($\rho_{sr}$), a total weighted average based on the composition of rock type for the entire area ($\rho_{avg}$), and an azimuthally dependent weighted average based on the rock type in each of the four Cartesian quadrants ($\rho_{quad}$). The values for these densities can be seen in \cref{tab:rock_density}.

\begin{table}
    \centering
    \renewcommand*{\arraystretch}{1.1}
    \setlength{\tabcolsep}{0.9em}
    \begin{tabular}{c | c}
       Density Profile &  Density (\unit[per-mode = symbol]{\gram\per\cubic\centi\metre})\\
       \hline
        $\rho_{sr}$ & 2.650 \\
        $\rho_{avg}$ & 2.768\\
        \hline
        $\rho_{quad}$: Q1 & 2.765\\
        $\rho_{quad}$: Q2 & 2.802\\
        $\rho_{quad}$: Q3 & 2.806\\
        $\rho_{quad}$: Q4 & 2.752 \\
    \end{tabular}
    \caption{Rock density values used to evaluate systematic uncertainties from rock composition.}
    \label{tab:rock_density}
\end{table}

The total underground muon fluxes between the 9 mountain profiles were compared using MUTE simulations. The fluxes resulting from the $\rho_{avg}$ and $\rho_{quad}$ profiles differed by less than 1\,\% for all sites. The fluxes from the $\rho_{sr}$  differed by less than 3\,\% compared to the $\rho_{avg}$ and $\rho_{quad}$ cases. Based on the USGS and Colorado geologic survey data and the relatively small differences seen between the density profiles, the average density profile $\rho_{avg}$ is used for full-scale simulations, and a 3\,\% systematic uncertainty is assigned to account for rock inhomogeneities. 

\subsubsection{Air Gaps}\label{sec:airgaps}
Due to the complex topology of the mountainous region surrounding the EEM, some $(\theta, \phi)$ directions may include air gaps as illustrated in blue in \cref{fig:airgap}. To include the air gaps, the zenith and azimuthal angles of each site mountain profile were projected onto a narrow column, and the full 2$\pi$ steradian was swept to check for repeating $R$ values. No air gap exists if only a single $R$ value is found. 
If an odd number of $R$ values is found, one or more air gaps are present before the muon reaches the lab, and a correction is needed. If an even number of $R$ values is found, an air gap is also present, but the muon entered the world volume underground and has, at minimum, 10s of km.w.e of shielding, and $R$ is set to a maximum value $R_{max}$.
From this reasoning, based on the number, $N$, of repeating $R$ values found, a corrected total $R$ value, $R_{corr}$, was calculated as:


\begin{figure}
    \centering
    \includegraphics[width = 8.5cm]{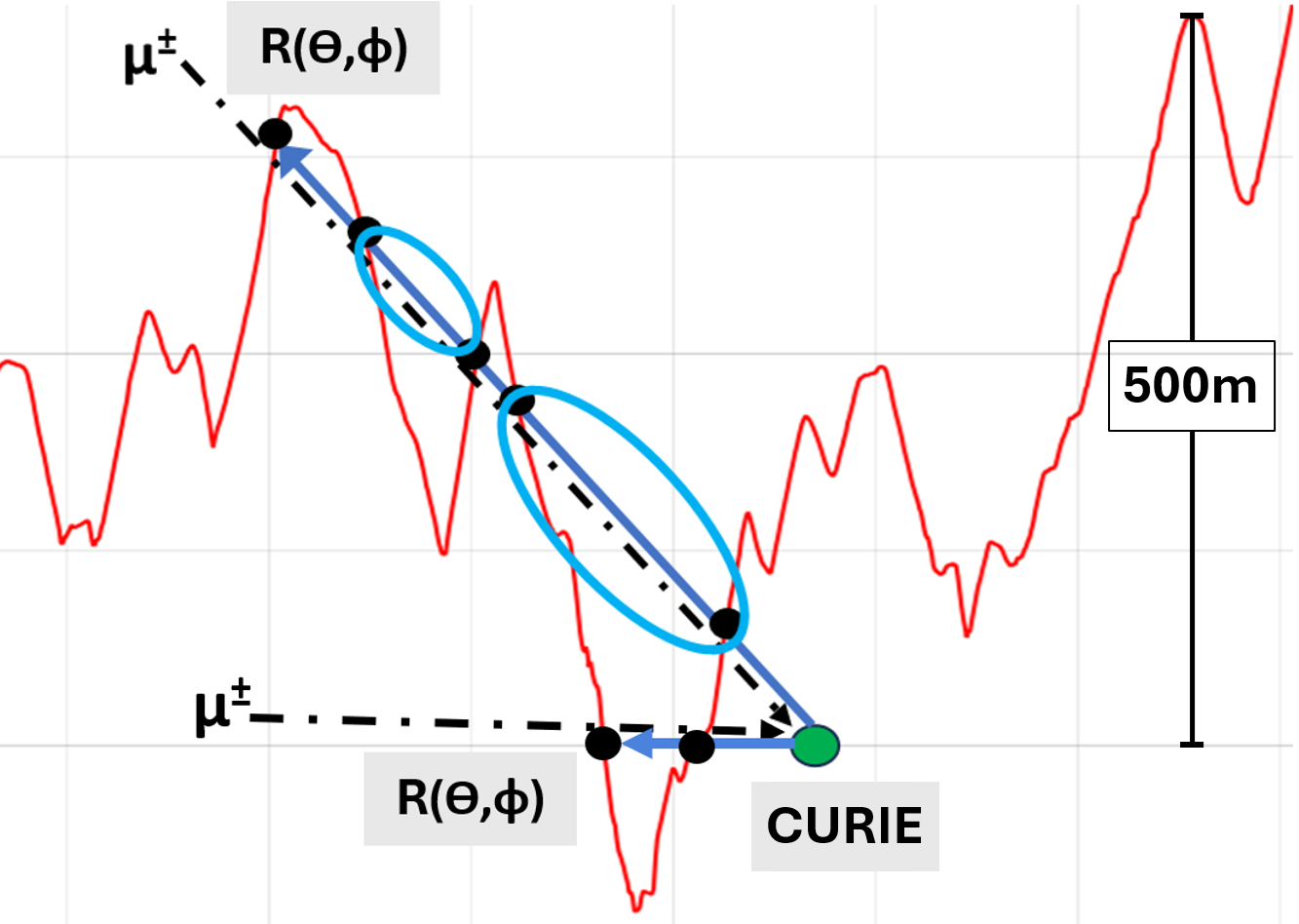}
    \caption{Cross-section elevation plot showing an instance of an odd repeated $R$ (upper trajectory) and even repeated R (lower trajectory) for the given ($\theta, \phi$), which would lead to an overestimation of $\Chi$, as described in the text.}
    \label{fig:airgap}
\end{figure}


\begin{equation}
    R_{\text{corr}} = 
    \begin{cases} 
        R, & \text{if } N = 1 \\ 
        R_{\text{max}}, & \text{if } N \text{ is even} \\ 
        R_{1} + \sum\limits_{i=1}^{\frac{N-1}{2}} (R_{2i+1} - R_{2i}), & \text{if } N \text{ is odd}
    \end{cases}
\end{equation}

The above process was carried out for \ang{0.01}, \ang{0.1}, \ang{0.5}, and \ang{1} ($\theta, \phi$) sampling resolutions. Simulations were run using the corrected mountain profiles for each resolution and compared to the uncorrected mountain profiles. The corrections for all projection resolutions yielded total underground fluxes that were between 1 and 2\,\% higher than the uncorrected mountain profile. Based on these results, a correction resolution of \ang{1} was chosen to be sufficiently accurate while being computationally efficient, leading to an average increase in the flux of 1.5\,\%. 

\subsection{Surface muon flux}

\subsubsection{Seasonal Variations}\label{sec:seasonalfluxes}
The effect of seasonal modulations in the muon flux is a well-known phenomenon~\cite{borexino, CRU}. The atmosphere warms and expands during the summer, decreasing the atmospheric density. Pions and kaons, the primary producers of atmospheric muons, are less likely to interact with the atmosphere, resulting in an increased probability of directly decaying to high-energy muons lower in the atmosphere. This process is the opposite in the winter months, and thus, the muon flux at the surface is lower in the summer months but higher underground due to the higher mean muon energy.

\begin{figure}
    \centering
    \includegraphics[width = 8.5cm]{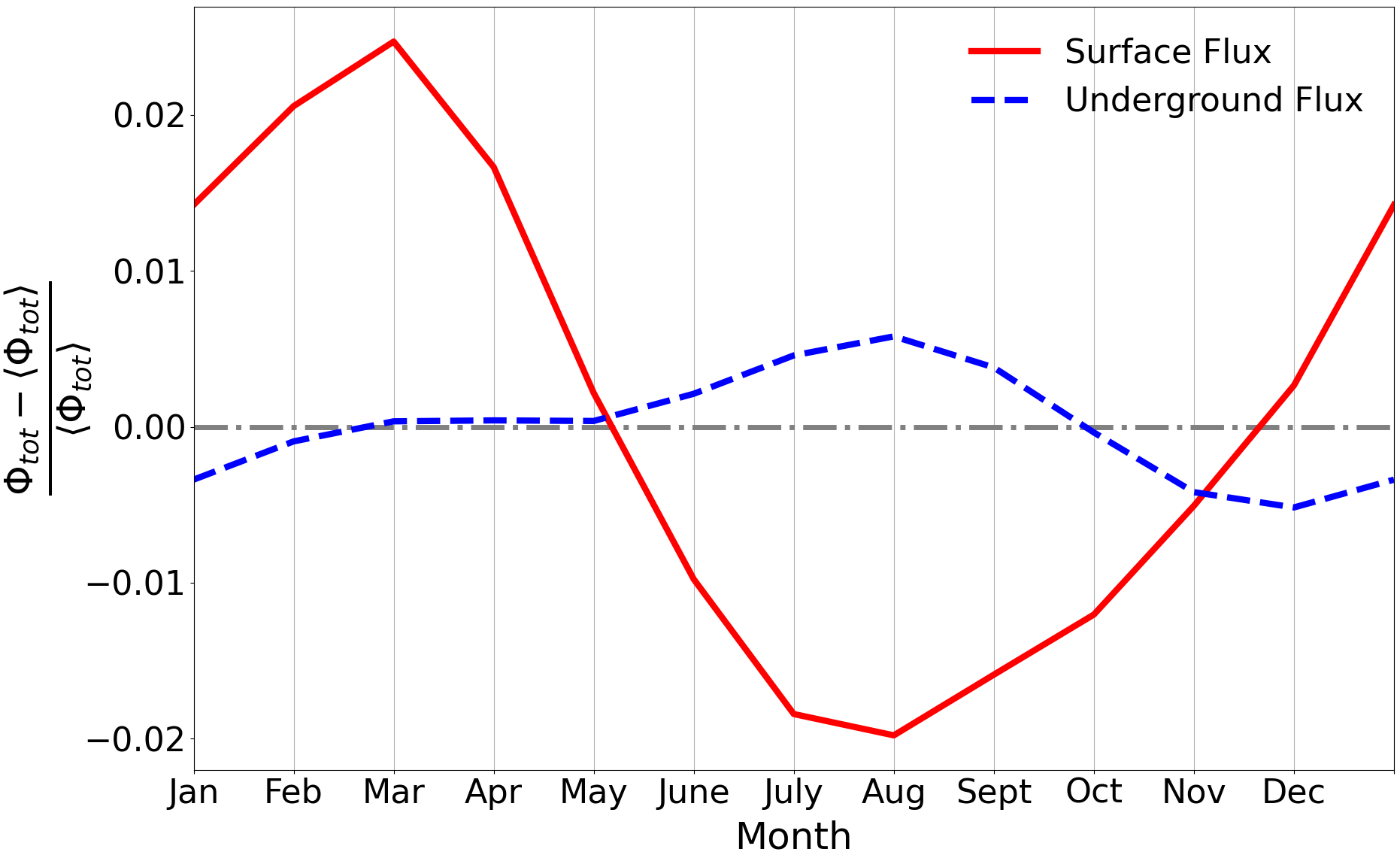}
    \caption{Estimated seasonal surface and underground muon flux variations over a year for the EEM. The fluxes are normalized to the yearly average to better show trends.}
    \label{fig:eem_season}
\end{figure}

To estimate the seasonal variations, we used MUTE with input from the NRLMSISE-00 atmospheric model~\cite{msis}. In this instance, Daemonflux is not used for the surface fluxes, but MCEq is instead used directly to specify a geographic location. The seasonal surface and underground fluxes for the EEM can be seen in \cref{fig:eem_season}. The interaction model used for these calculations was SIBYLL-2.3c~\cite{sibyll2.3c}. Current high-energy hadronic interaction models underestimate low to mid-energy muon abundances in high-energy cosmic ray air-showers and, therefore, also underestimate the muon flux at shallower depths~\cite{Coleman:2022abf, woodley2024cosmicraymuonslaboratories}. However, since we are taking the ratio of monthly surface and underground fluxes to their respective yearly averages, this underestimation is expected to be largely averaged out, leaving only the expected yearly trend.

\subsubsection{Altitude Variations}\label{sec:altitude}
\begin{figure}
    \centering
    \includegraphics[width=8.5cm]{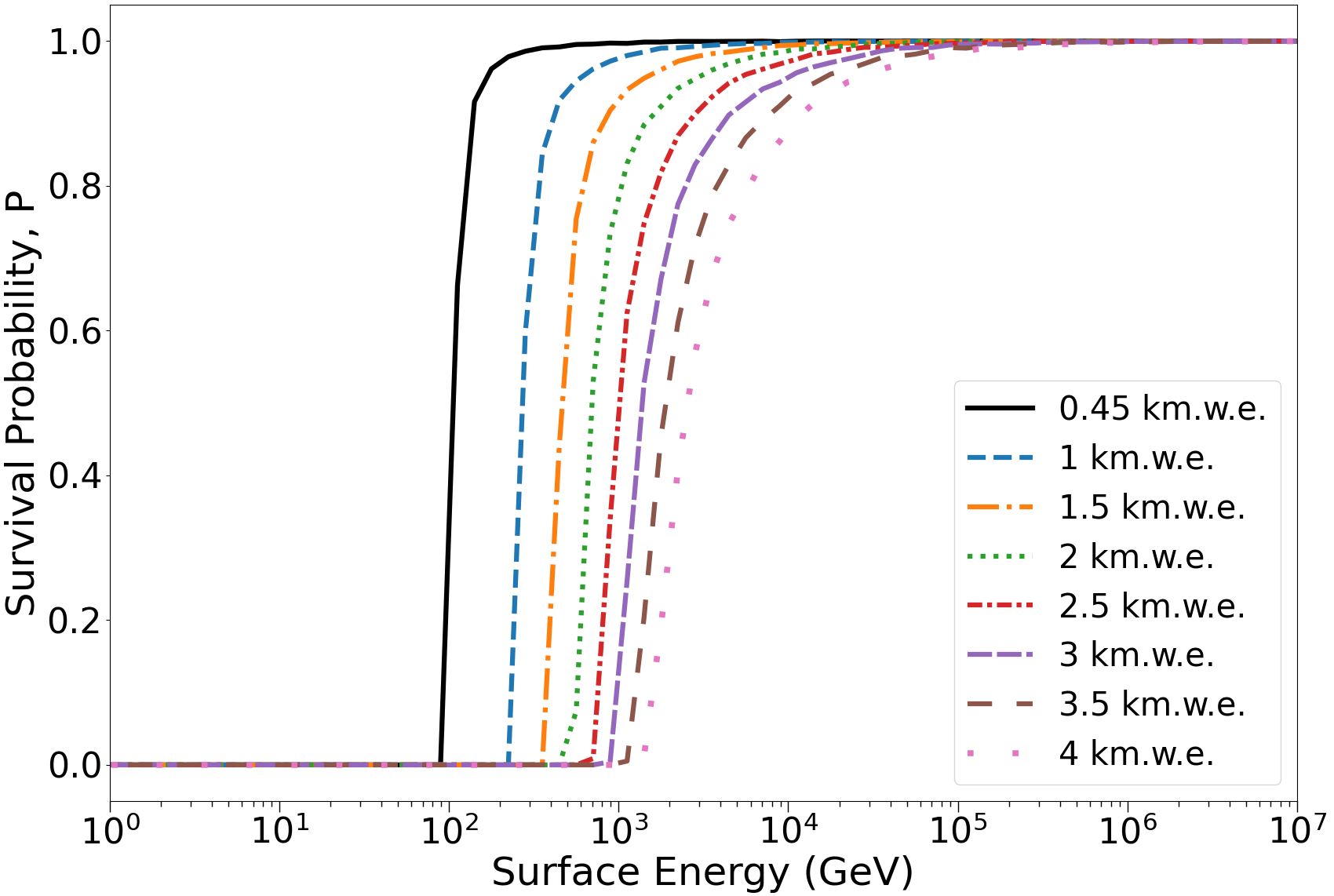}
    \caption{Simulated survival probability of muons for various slant depths at Site 2. The 0.45 km.w.e. line corresponds to a depth shallower than the minimum calculated depth between all three sites.}
    \label{fig:surface_surv_prob}
\end{figure}

The effect of the altitude on the muon surface flux has been studied in great detail \cite{mu_atmo_flux, mu_atmo_flux2,  mu_atmo_flux3,  mu_atmo_flux4,  mu_atmo_flux5,  mu_atmo_flux6} and, in general, the muon flux increases with altitude up to $\approx$ \SI{10}{\km} above sea level. For most shallow and deep underground labs, using the flux at sea level for the initial propagation is sufficient in predicting the underground fluxes, for depths $\gtrsim$ 0.5 km.w.e. In the case of the EEM, with an altitude of \SI{2.4}{\km} above sea level, only lower energy muons below $\approx$ \SI{20}{\GeV} are added to the surface spectrum. The calculated minimum slant depth from the mountain profiles for Site 0, 1, and 2 are 0.497, 0.485, and 0.467 km.w.e, respectively. Thus, only incident muons above $\approx$ \SI{90}{\GeV} should penetrate to the depths of our locations \cite{mu_vs_height}. To study this, we compared the survival probabilities from Site 2 for slant depths of [0.45, 4.0] km.w.e. vs. incident muon surface energy, using Daemonflux and MUTE (see \cref{fig:surface_surv_prob}), and find no contribution to the underground flux below $\approx$ \SI{90}{\GeV} surface energy, in agreement with previous simulation work \cite{mute}. Thus, we use the sea level surface flux model in our simulations, and compare our results to other underground facilities at both similar or lower altitudes.

\subsection{Underground Flux Simulations}\label{sec:CompTools}
\begin{figure}[!b]
    \centering
    \includegraphics[width = 8.5cm]{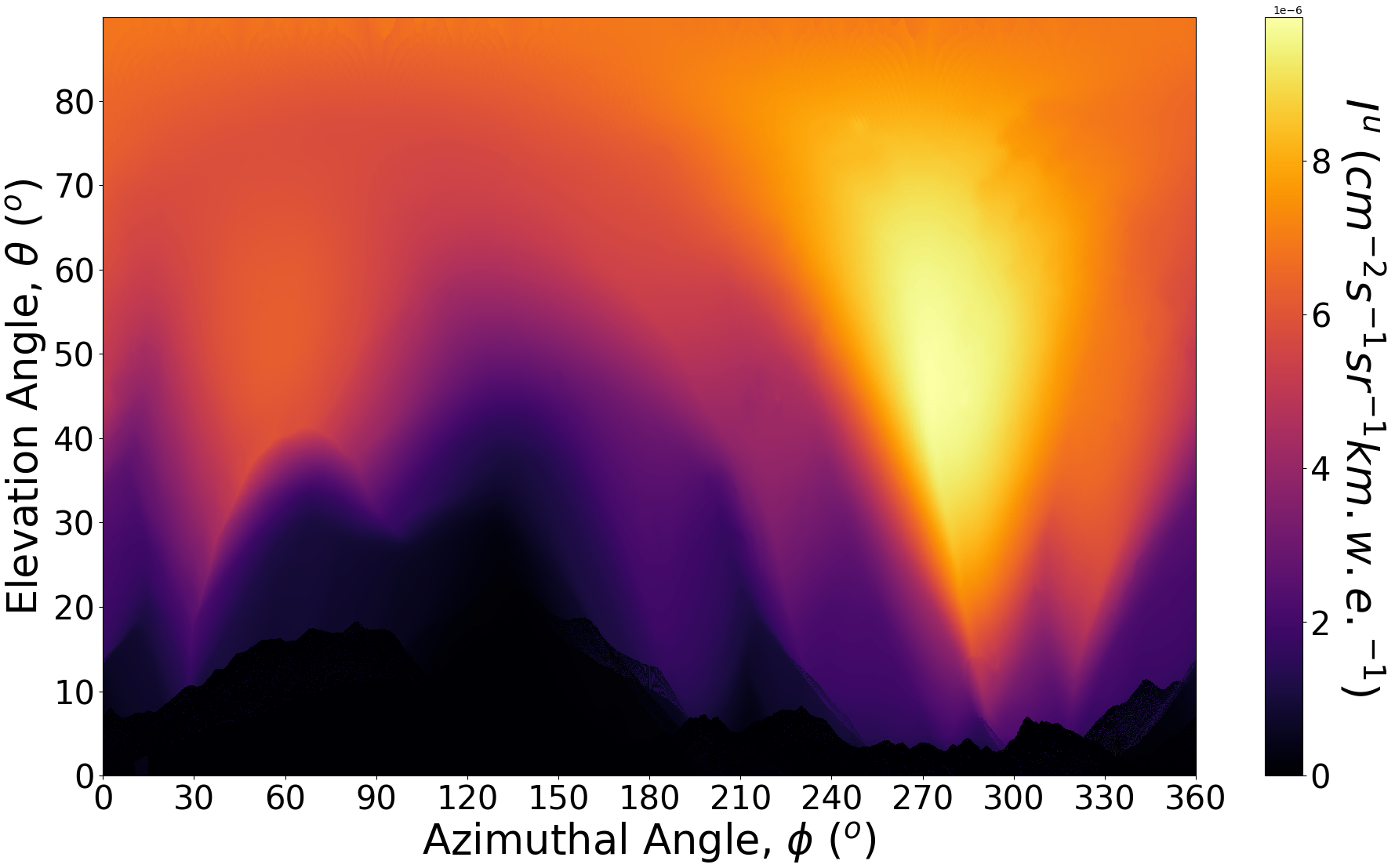}
    \caption{Heatmap of the underground muon intensities as a function of arrival angle for Site 1. Note that the y-axis is given as an elevation angle for a more straightforward visual interpretation.}
    \label{fig:heatmap}
\end{figure}
\begin{figure}[!t]
    \centering
    \includegraphics[width = 8.5cm]{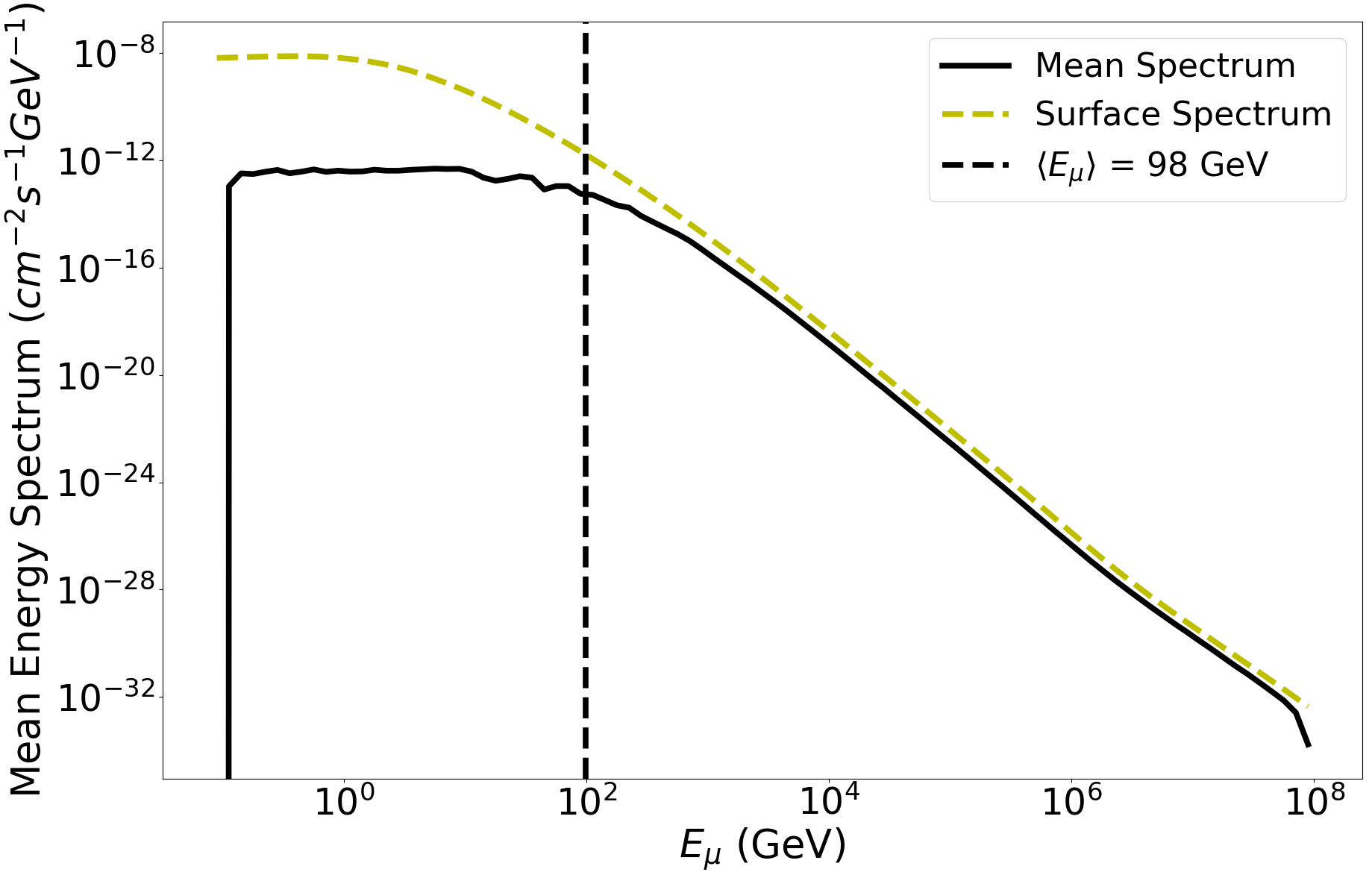}
    \caption{The mean underground muon energy spectrum between all sites, shown relative to the surface spectrum. The vertical line indicates the mean underground muon energy.}
    \label{fig:MuonSpectrum}
\end{figure}

The full-scale muon flux simulations were performed for sites 0, 1, and 2 using the $\approx$ \SI{84}{\km^{2}} area surrounding the EEM (detailed in \cref{sec:mtnprofiles}) using the density and air gap corrections described in \cref{sec:rock} and \cref{sec:airgaps}. For the surface flux modeling, the U.S. Standard Atmosphere is used~\cite{us_atmo}. In the simulations, just over 30 million total muons were propagated. The results yielded fluxes of $\phi_{\mu}$ = 0.227 $\pm$ 0.023, 0.217 $\pm$ 0.022, and 0.259 $\pm$ 0.026 $\mu\text{/}$\unit[per-mode = symbol]{\square\metre\per\second} for sites 0, 1, and 2, respectively. Due to the large statistics, the reported uncertainties in the fluxes are purely systematic, representing the uncertainties from the Global Spline Fit (GSF) and Data-Driven (DDM) models and the uncertainties from the expected geological systematics, described earlier. The arrival direction-dependent heat map of muon intensities is shown in \cref{fig:heatmap} for Site 1. Additionally, we used MUTE to calculate the expected underground muon energy spectrum and the mean muon energy for each Site. The spectra and mean energy between all sites are nearly identical, therefore, in Figure \ref{fig:MuonSpectrum} we show the expected mean underground muon energy spectrum. This is shown relative to the surface energy spectrum, along with the mean muon energy of $\langle E_{\mu}\rangle \text{ = }$ \SI{98}{} $\pm$ \SI{2}{\GeV}.
\subsubsection{Depth-Intensity Relation}\label{sec:model}
A convenient way of comparing underground research facilities with different overburden rock profiles is to normalize them to a flat overburden. Doing so allows for the direct comparison of cosmogenic background reduction between facilities under mountains and those with flat overburdens. To accomplish this conversion, a depth-to-intensity relationship is needed.

For deep underground facilities, those with overburdens $\geq$ 1 km.w.e., a Depth-Intensity Relation (DIR) was developed by Mei and Hime (M$\&$H) \cite{mei_hime}. This DIR, which spans 1 to 10 km.w.e., has been the standard for nearly 2 decades, but for those facilities with overburdens $<$ 1 km.w.e., it must extrapolate to these depths. Another DIR, which spans 0 to 6 km.w.e.\ and includes shallow depths, was developed by Mitrica et al.\ \cite{mitrica}. This model, however, seems to over-predict the km.w.e.\ shielding when compared to world data.
Parametric models which are typically used to develop a DIR do so by fitting the vertical-equivalent underground muon intensities, for instance, in Refs. \cite{mei_hime, macro, sudbury, lvd}. While it has been shown for depths of $\leq$ 0.1 km.w.e.\ the function relating the integrated and vertical-equivalent muon intensities is nearly constant \cite{Bogdanova_2006}, in general, the vertical-equivalent intensities are known to only approximate the true vertical-intensities for $\theta \lesssim$ \ang{30} \cite{barrett, Bugaev}. Therefore, we present a new DIR to predict the equivalent depth relative to a flat overburden, by modeling the total integrated underground flux through a large range of depths.

\begin{figure}
    \centering
    \includegraphics[width = 8.5cm]{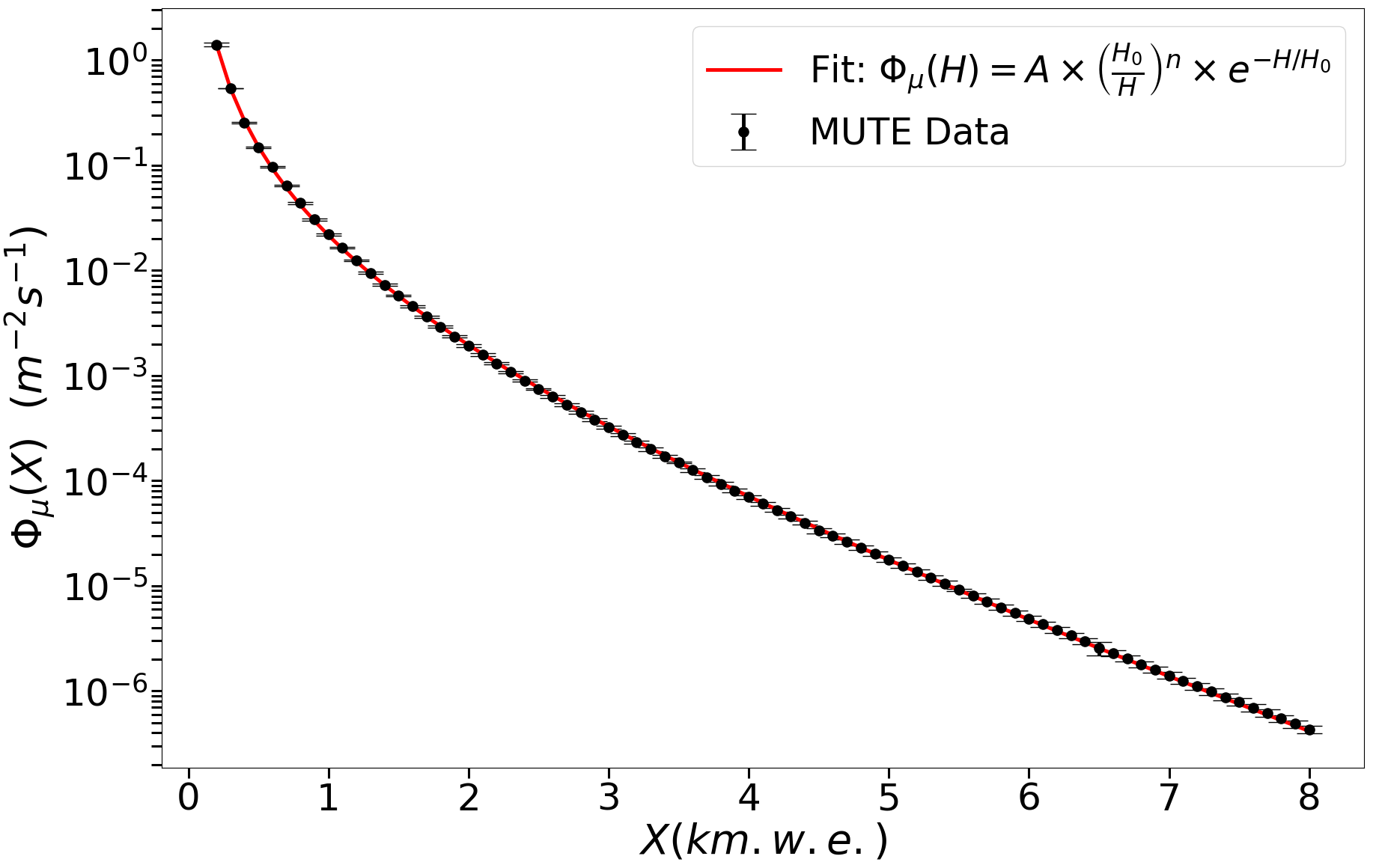}
    \caption{Fit function applied to the simulated underground muon flux data for flat overburden depths of 0.2 - 8.0 km.w.e.}
    \label{fig:model_fit}
\end{figure}

To achieve this, Daemonflux and MUTE were used to run simulations for a flat overburden profile for depths of 0.2 - 8.0 km.w.e., using standard rock. A fit function, which follows the depth intensity relation given in Ref.\ \cite{lipari}, was then defined to model the data (shown in \cref{fig:model_fit}):

\begin{equation}
    \Phi_{\mu}(H) = A\times \left(\frac{\Chi_{0}}{\Chi}\right)^{n}\times e^{-\Chi/\Chi_{0}}\text{.}
    \label{eq:kmwe_model}
\end{equation}

\begin{figure}
    \centering
    \includegraphics[width = 8.5cm]{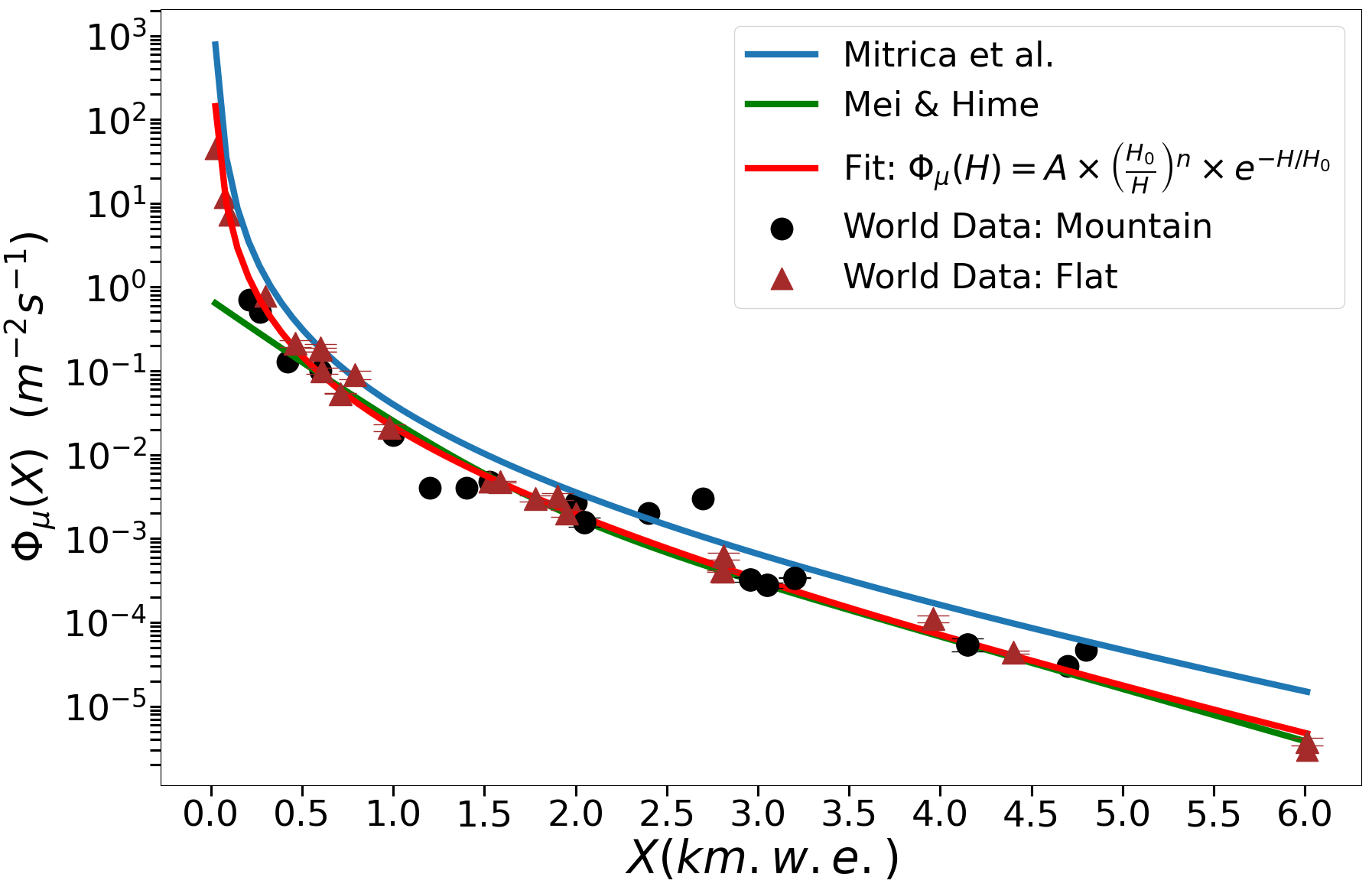}
    \caption{M$\&$H and Mitrica et al.\ flux models compared to $\Phi_{\mu}(H)$, plotted against the experimental data from worldwide facilities taken from Refs.\ \cite{mei_hime, mitrica,oroville, braidwood, chooz, callio}. Error bars on the experimental data are shown where available.}
    \label{fig:model_comparison}
\end{figure}

\begin{figure}
    \centering
    \includegraphics[width = 8.5cm]{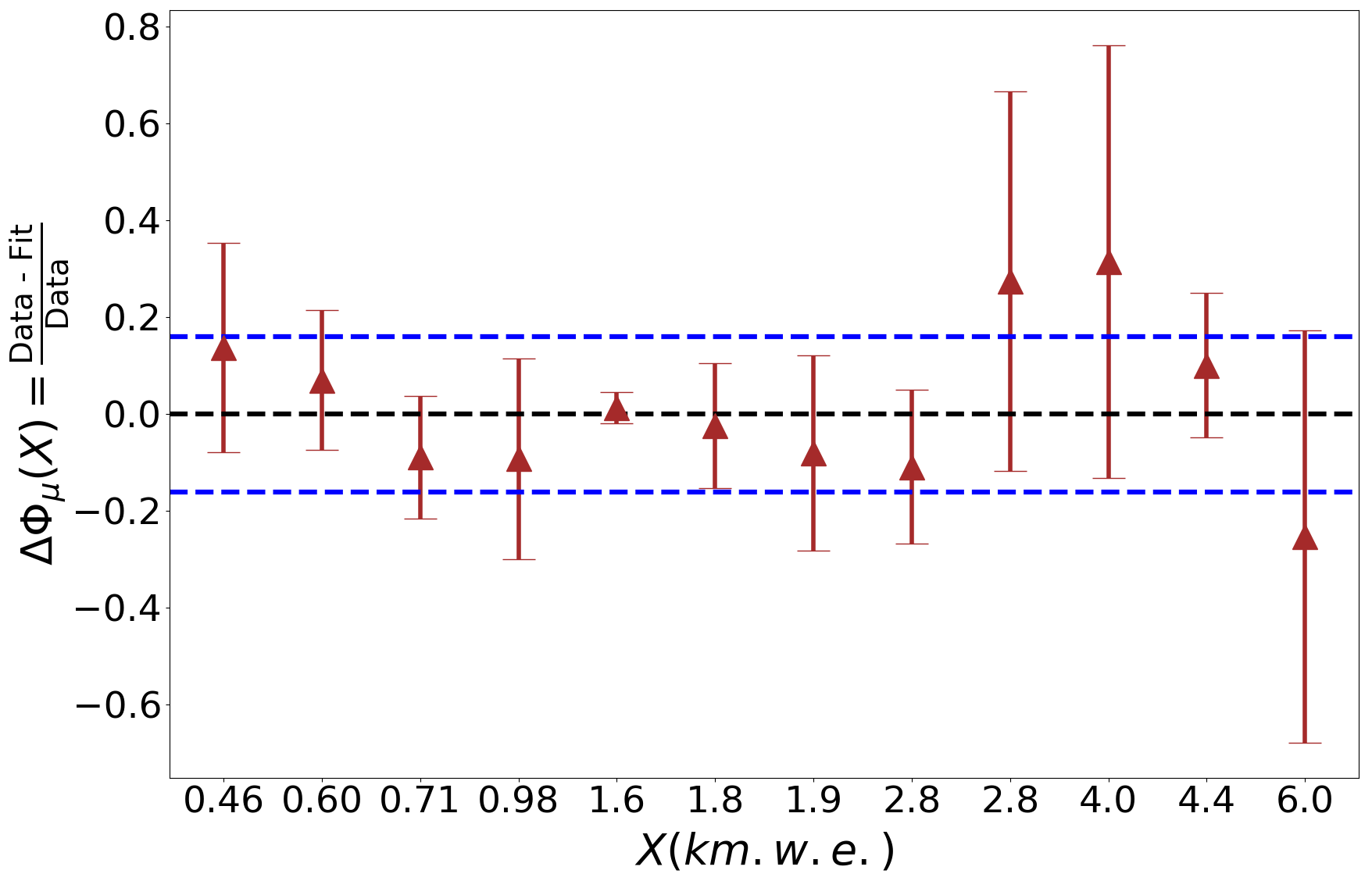}
    \caption{The residuals between the flat overburden World data and $\Phi_{\mu}(H)$. The blue horizontal lines indicate the root-mean-square deviation of the residuals from the experimental uncertainties in the measurements.}
    \label{fig:flat_model_comparison}
\end{figure}

The parameters are $A = 0.044\text{ } \mu\text{/}$\unit[per-mode = symbol]{\square\metre\per\second}, $n = 2.145$ and is a unitless parameter, $H_{0} = 1.094$ km.w.e., $H$ is the equivalent depth in units of km.w.e., and $\Phi_{\mu}(H)$ is the total underground muon flux in units of $\mu\text{/}$\unit[per-mode = symbol]{\square\metre\per\second}. For facilities with flat overburdens, the equivalent depth is the true depth. For labs with mountain overburdens, the equivalent depth is the depth that the lab would be at for a flat overburden, normalized to the total underground muon flux. A comparison between $\Phi_{\mu}(H)$, M$\&$H, and the Mitrica models can be seen in \cref{fig:model_comparison}, where they are plotted against the experimental data taken from Table I in Ref. \cite{mei_hime} and Table II from Ref.\ \cite{mitrica}, as well as from Refs.\ \cite{oroville, braidwood, chooz, callio}. The data from these tables is a mix of flat and mountain overburden facilities. \cref{fig:flat_model_comparison} shows the relative difference between the experimental data and \cref{eq:kmwe_model}, for a subset of facilities with a flat overburden.

\subsection{Simulation Results}\label{sec:simresults}
As stated earlier, the expected underground cosmic-ray muon flux, $\phi_{\mu}$, was simulated for Site 0, 1, and 2. The results yielded fluxes of $\phi_{\mu}$ = 0.227 $\pm$ 0.023, 0.217 $\pm$ 0.022, and 0.259 $\pm$ 0.026 $\mu\text{/}$\unit[per-mode = symbol]{\square\metre\per\second} for sites 0, 1, and 2, respectively. The new model presented in \cref{eq:kmwe_model}, was then used to convert these fluxes to the equivalent km.w.e.\ depth relative to a flat overburden. This procedure resulted in depths of 0.426 $\pm$ 0.014, 0.434 $\pm$ 0.014, and 0.405 $\pm$ 0.013 km.w.e.\ for Site 0, 1, and 2, respectively. The Mei and Hime and Mitrica et al.\ models were also used to predict the equivalent depth for each simulated $\phi_{\mu}$ value.


\section{Experimental Methods}\label{sec:ExpMethods}

\subsection{Detector Configuration and Efficiency}
\begin{figure}
    \centering
    \includegraphics[width=8.5cm]{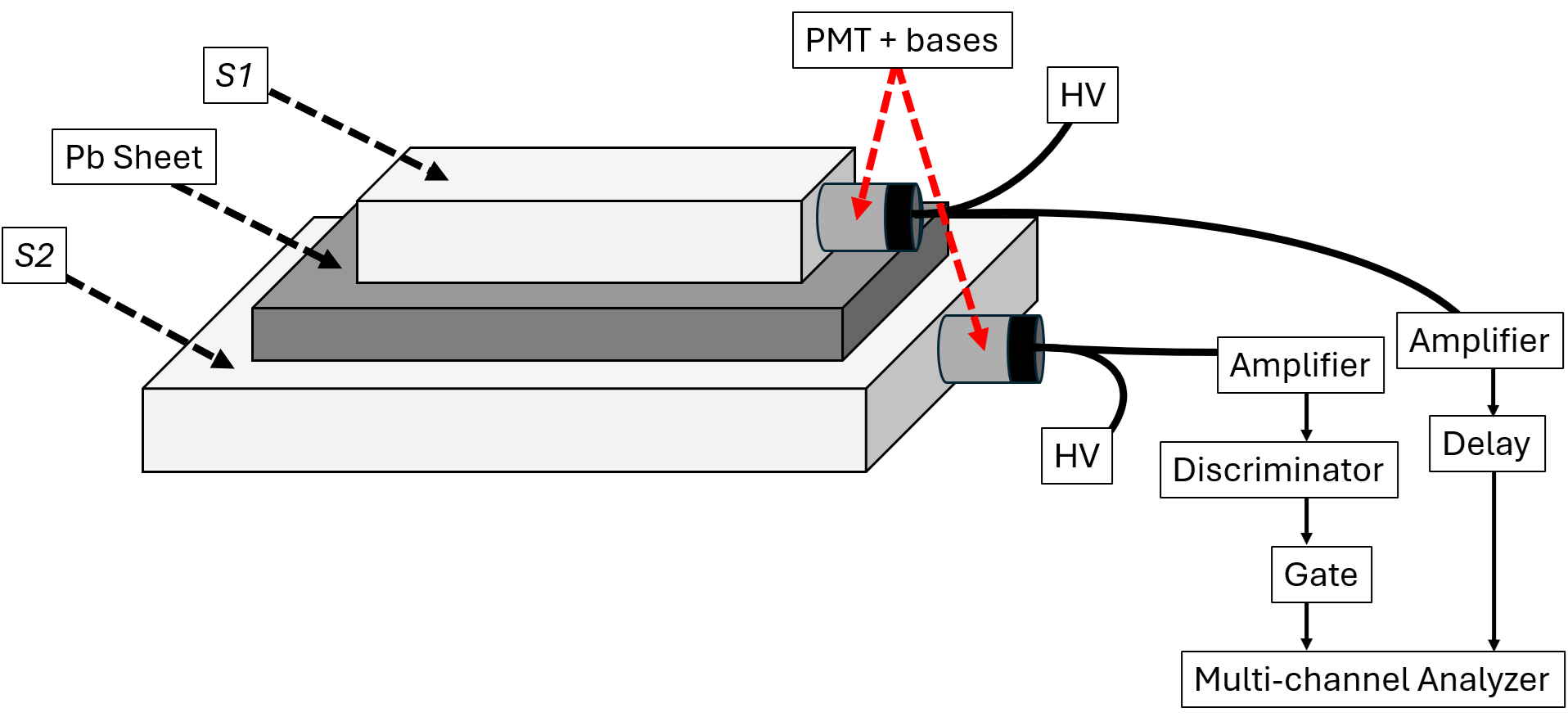}
    \caption{Schematic of the experimental apparatus highlighting scintillators $S1$ and $S1$, the lead sheet, the PMTs, and the block diagram of the main electronic components.}
    \label{fig:detector_schematic}
\end{figure}

To directly compare with simulation, the measurements of the total underground muon flux were performed in Site 0 and Site 1. The experimental setup consisted of two stacked, commercial plastic scintillators with a lead sheet separating the detectors. The top scintillator, $S1$, has dimensions \SI{12.5}{\cm} length $\times$ \SI{14.5}{\cm} width $\times$ \SI{4.0}{\cm} thickness. The bottom scintillator, $S2$, has dimensions \SI{30.0}{\cm} length $\times$ \SI{48.0}{\cm} width $\times$ \SI{4.0}{\cm} thickness. An uncertainty of $\pm$ \SI{0.5}{\cm} was assigned for each dimension. The lead sheet has dimensions \SI{18.0}{\cm} length $\times$ \SI{30.0}{\cm} width $\times$ \SI{2.0}{\cm} thickness. A schematic diagram of the setup can be seen in \cref{fig:detector_schematic}. The scintillators, $S1$ and $S2$, were coupled to individual photomultiplier tubes (PMT) and bases, each with an applied bias voltage of \SI{800}{\V}.

Without proper shielding, a dominant background source in underground facilities is the gamma-ray background originating from the alpha-decay chain of primordial isotopes \cite{eisenbud1997environmental}. While most of the muon energy deposited in the upper scintillator is above the gamma energies, to further discriminate the gamma background events against the cosmic muon events, the panels were run in coincidence mode. The coincidence mode employed here was set to a time window of $\approx$ \SI{4}{\micro\second}. A discriminator threshold corresponding to $\approx$ \SI{2}{\MeV} was used to predominately select muons in the larger panel. The experimental configuration was operated for a total live time of \SI{385.5}{\hour} and \SI{619.7}{\hour} in Site 0 and Site 1, respectively.

To calculate the detection efficiency of coincidence muon events, the top panel was also run in single mode. A channel cut, corresponding to all channels well above the distinguishable gamma background, was selected and a count rate was calculated using the number of observed counts and the detector live time. A count rate for the same channel cut of the coincidence data was also calculated. The ratio of the coincidence to single count rates was taken to give the total detector efficiency, $\epsilon \text{ = } R_{coin}/R_{single}$. This procedure was done for Site 0 and Site 1 which yielded efficiencies of $\epsilon_{\text{Site 0}} =$ 0.823 $\pm$ 0.039 and $\epsilon_{\text{Site 1}} =$ 0.799 $\pm$ 0.035. The coincidence spectra was analyzed for the values presented later and for the comparison of the measured energy spectrum with simulation.


\subsection{Measurement Corrections}\label{sec:MeasCorrections}

\subsubsection{Geometric Corrections}\label{sec:GeoCorrections}
Due to the complex nature of the underground muon distribution and the geometry of our experimental setup, we must estimate the effective detector area. This was achieved through Monte Carlo simulations using Geant4~\cite{g4}. Our detector configuration was reconstructed in Geant4, inside of a \SI{6.6}{\m} $\times$ \SI{6.6}{\m} $\times$ \SI{0.21}{\m} rectangular volume. The volume was chosen such that a particle produced in any of the upper four corners could pass directly between the top and bottom detectors, only interacting with the lead sheet, in order to probe the entire trajectory space. As this was a purely geometric simulation, geantinos (particles that do not interact with anything) were chosen as the ``particle" for the simulations, and no physical interactions were necessary. The geantinos were produced from the upper planar surface ($z = \SI{0.21}{\m}$) of the volume with a randomly sampled ($x$, $y$) start position. The geantinos were propagated downward with a momentum direction randomly sampled from a cosine distribution to produce an isotropic flux~\cite{cosine_law}. After this procedure, good agreement was found between thrown geometries and an isotropic distribution. Any geantino that passed through both the top and bottom detector was recorded as a hit, mimicking the coincidence trigger of the real detector.


Since the underground muon distribution is not isotropic, each combination of $\theta$ and $\phi$ value was recorded for all geantinos propagated and separately for those that triggered a coincidence. These angles were then compared to the MUTE simulation angles and were weighted by the simulated underground muon intensities from MUTE to reproduce the expected underground muon distribution. A heat map of the underground muon intensities from the Geant4 weighting procedure was produced and compared to the $2\pi$ steradian heat map directly from MUTE, where the zenith-dependent detector response is observed. Additionally, simulations were run for a change in the detector dimensions by $\pm$ \SI{0.5}{\cm}, to account for the error in the measurements of the dimensions.

The effective area is given by,  

\begin{equation}
    A_{eff} = \frac{I_{coin}}{I_{tot}}\times A_{gen}\text{,}
\end{equation}

where $I_{coin}$ and $I_{tot}$ are the sums of the intensities over all angles for coincidence triggers and all geantinos, respectively, and $A_{gen}$ is the generation area of the \SI{6.6}{\m} $\times$ \SI{6.6}{\m} planar surface. The results of these simulations estimated a value of $A_{eff}$ = 241.15 $\pm$ 23 (sys.) $\pm$ 3.24 (stat.)\,\unit{\square\centi\metre}. Lastly, we note that the contribution of potential coincidence triggers coming from the walls of the volume was tested using an \SI{18}{\m} $\times$ \SI{18}{\m} $\times$ \SI{0.21}{\m} volume and was shown to be negligible.


\subsubsection{Gamma Background}\label{sec:GammaCorr}
\begin{figure}[!htb]
    \centering
    \includegraphics[width=8.5cm]{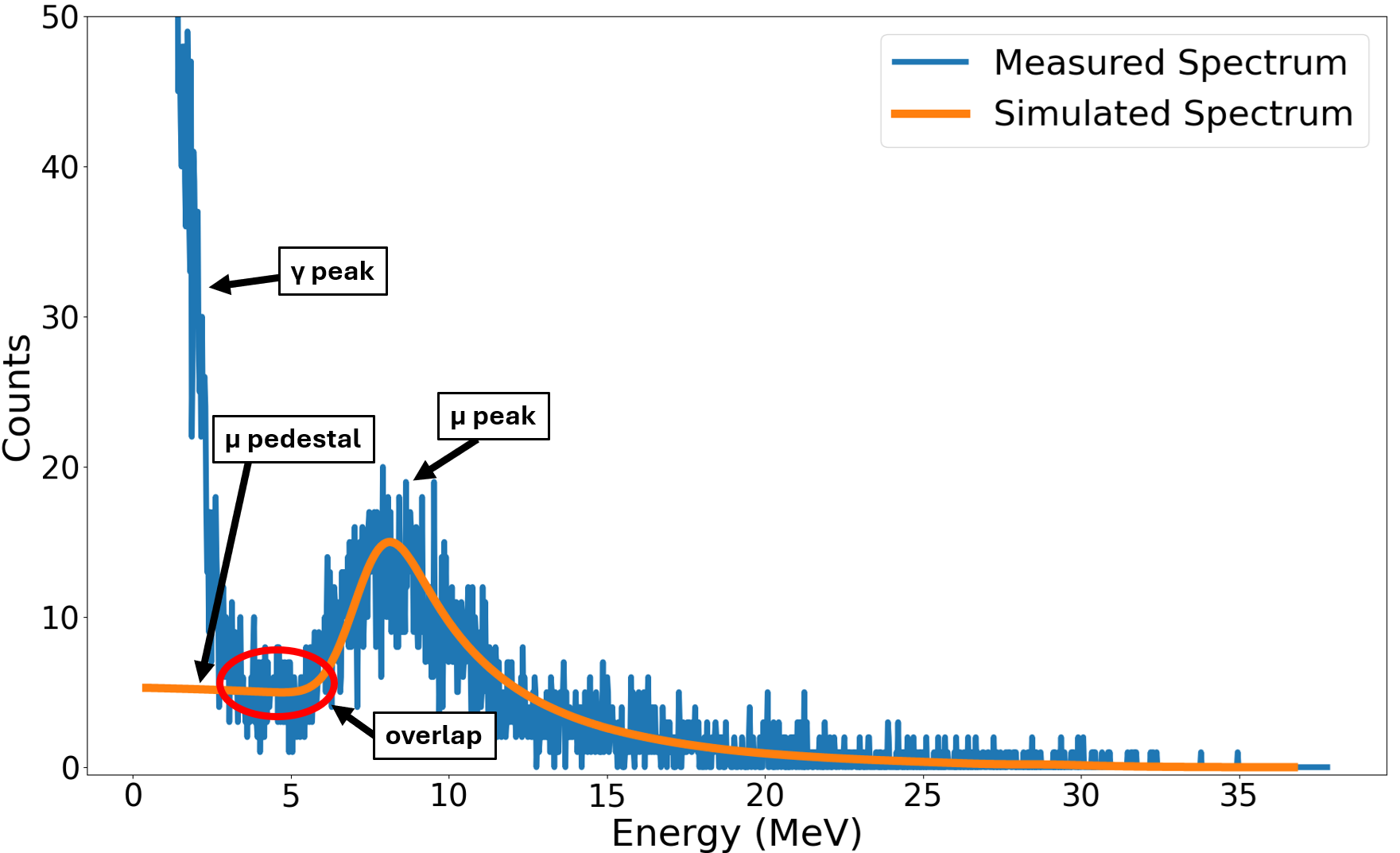}
    \caption{Coincidence scintillator energy spectrum from Site 1 highlighting the gamma and muon peaks, the overlap of the tails, and the expected muon pedestal. The simulated spectrum is also shown, where a muon energy loss of \qty[per-mode = symbol]{2}{\MeV\per\centi\metre} was applied to the simulation track lengths.}
    \label{fig:site1_full_spectrum}
\end{figure}

Whether on the surface or underground, gamma rays are present from naturally occurring sources, such as $^{40}$K and the U and Th decay chains. These gamma rays contribute to the measured background noise in many experiments. While they can be shielded using high-Z materials, in general, they must be accounted for in most low-background experiments. In the experimental setup used here, this gamma background manifests itself in the form of low-energy (or low-channel) counts, typically below \SI{3}{\MeV} of deposited energy. This obscures the total number of muons detected in the low energy region, which is expected to be an approximately constant muon count, the so-called muon pedestal. A channel-to-energy conversion was achieved by converting the extracted geantino track length spectrum, from the simulations detailed in \cref{sec:MeasCorrections}, to an energy spectrum via an ionization energy loss of \qty[per-mode = symbol]{2}{\MeV\per\centi\metre}. This corresponds to a minimum ionizing energy loss for muons in a typical plastic scintillator ~\cite{mips}. The measured and simulated spectra were peak-matched to extract a channel-to-energy conversion factor, and the measured spectrum was converted to an energy spectrum to identify physical features. \cref{fig:site1_full_spectrum} shows the converted measured coincidence spectrum from Site 1 and the simulated spectrum, highlighting the main features. The simulated spectrum is normalized to the experimental data.

To estimate the gamma background, an exponential was fit to the data for low channels, in the form of,

\begin{equation}
    y = a e^{-b x} + c\text{.}
    \label{eq:gammamodel}
\end{equation}

\begin{figure}
    \centering
    \includegraphics[width=8.5cm]{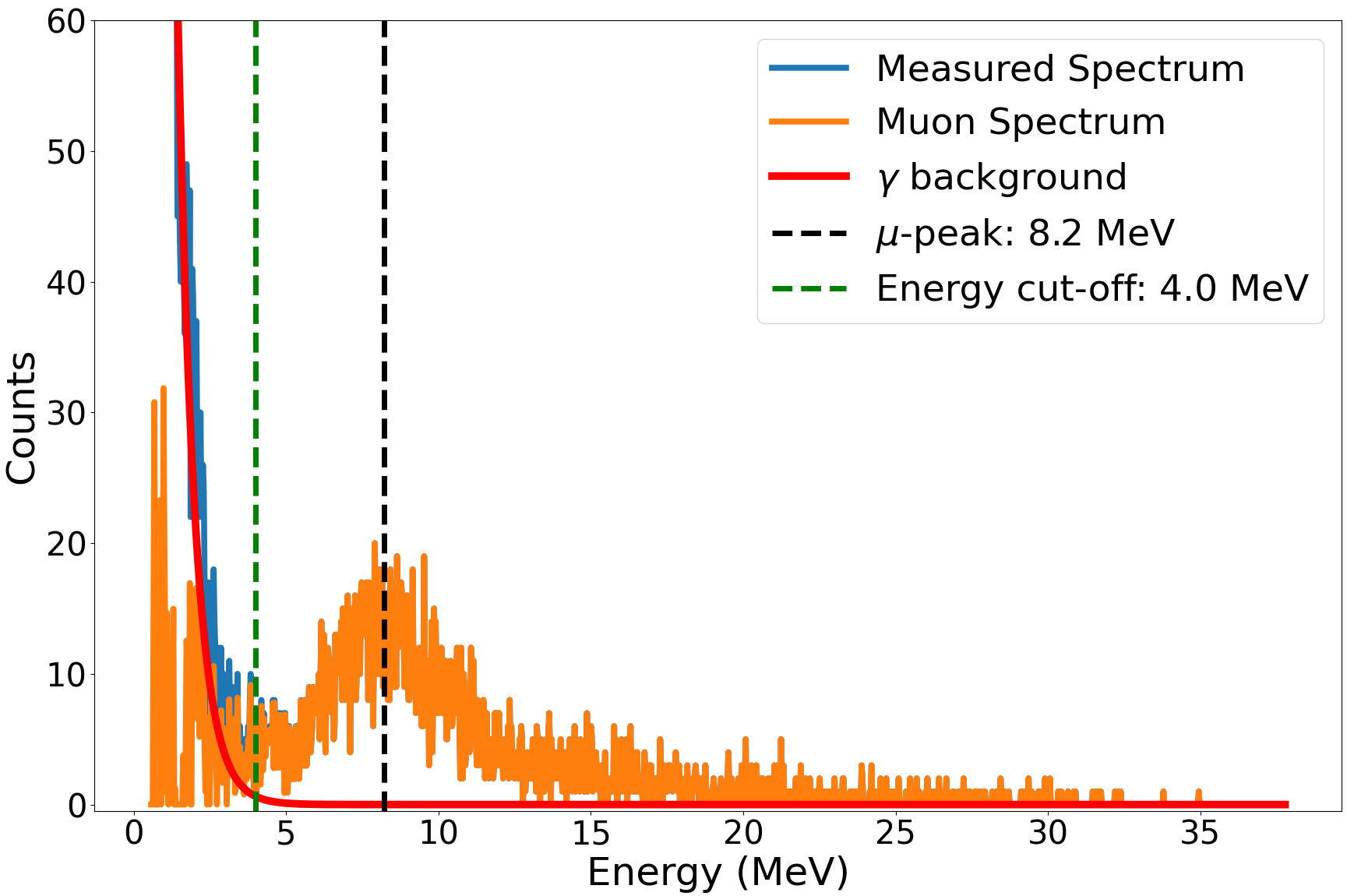}
    \caption{Results of the channel-to energy-conversion and $\gamma$ background subtraction from the measured spectrum in Site 1. The estimated energy cut-off, $\mu$-peak, and $\gamma$ background are also shown.}
    \label{fig:site1_mu_minus_gamma}
\end{figure}

In each of the spectra analyzed, the choice in the channel number upper range was an approximate midway point between the gamma and muon peaks. The expected contribution of counts from the gamma background toward higher channel number should go to zero, thus the offset in \cref{eq:gammamodel} represents the muon pedestal below the channel cut-off. Thus, we subtract off only the exponential term, preserving the muon counts for low energy channels, leaving the expected muon spectrum. The result of this procedure for Site 1 can be seen in \cref{fig:site1_mu_minus_gamma}, where we estimate an energy cut-off of \SI{4}{\MeV} for the fitting procedure and a muon peak of \SI{8.2}{\MeV}. After the gamma background subtraction, we arrive at an adjusted estimated total muon count of $N_{tot}$ = 6786 $\pm$ 60 and 10302 $\pm$ 230 for Site 0 and Site 1, respectively. The errors are systematic from the uncertainty in the fitting parameters.

\subsection{Experimental Results}\label{sec:ExpResults}
The underground muon flux ($\phi_{\mu}$) is calculated using the adjusted muon count ($N_{tot}$), the detector live time ($T_{live}$), $A_{gen}$, and the total efficiency of the detector ($\epsilon$) as

\begin{equation}
    \phi_{\mu} = \frac{N_{tot}}{T_{live}\times A_{gen}\times \epsilon}\text{.}
    \label{eq:flux}
\end{equation}

\vspace{0.5cm}
\begin{table*}[!htb]
    \footnotesize
    \centering
    \renewcommand*{\arraystretch}{1.1}
    \setlength{\tabcolsep}{0.6em}
        \begin{tabular}{l | c |ccc |c |ccc}
            & \multicolumn{4}{c|}{Measurement} & \multicolumn{4}{c}{Simulation} \\
            \hline
            \multirow{2}{*}{Location} & $\Phi$ Meas &$\Phi_{\mu}(H)$ & Mitrica et al. & M$\&$H & $\Phi$ Sim & $\Phi_{\mu}(H)$ & Mitrica et al. & M$\&$H \\
             & [$\mu\text{/}$\unit[per-mode = symbol]{\square\metre\per\second}] & \multicolumn{3}{c|}{$\Chi$ [km.w.e]} & [$\mu\text{/}$\unit[per-mode = symbol]{\square\metre\per\second}]  & \multicolumn{3}{c}{$\Chi$ [km.w.e]} \\
            \hline
            Site 0 & $0.246 \pm 0.020 \pm 0.012$ & $0.413 \pm 0.018$ & 0.546 & 0.306 & $0.227 \pm 0.023$ & $0.426 \pm 0.014$ & 0.562 & 0.329\\
            Site 1 & $0.239 \pm 0.025 \pm 0.010$ & $0.418 \pm 0.020$ & 0.552 & 0.314 & $0.217 \pm 0.022$ & $0.434 \pm 0.014$ & 0.570 & 0.342\\
            Site 2 & - & - & - & - & $0.259 \pm 0.026$ & $0.405 \pm 0.013$ & 0.536 & 0.290\\
            Sea Level & $\approx$  166.7 \cite{sea_level} & - & - & - & $172.71 \pm 1.72$ & - & - & - \\
        \end{tabular}
    \caption{Summary of found fluxes and depths. The experimentally measured fluxes and corresponding depths are reported, as well as the simulated fluxes and derived depths.}
    \label{tab:resultsummary}
\end{table*}

Based on our analysis, we find an underground muon flux for Site 0 of $\phi_{\mu} = 0.246 \pm 0.020_{sys.} \pm 0.012_{stat.}$ $\mu\text{/}$\unit[per-mode = symbol]{\square\metre\per\second} and for Site 1 a muon flux of $\phi_{\mu} = 0.239 \pm 0.025_{sys.} \pm 0.010_{stat.}$ $\mu\text{/}$\unit[per-mode = symbol]{\square\metre\per\second}. The overburden from these sites represents an approximate reduction by a factor of 700 compared to the flux at sea level. Using the measured values, we have reevaluated the predicted flat overburden km.w.e.\ overburden using \cref{eq:kmwe_model}, which yields 0.413 $\pm$ 0.018 km.w.e.\ and 0.418 $\pm$ 0.020 km.w.e., for Site 0 and Site 1 respectively. A summary of all of the muon flux models, as well as the results from the experimental measurements and simulations, can be found in \cref{tab:resultsummary}.


\subsection{Directional Flux Ratio Measurements}
The simulations from MUTE predict a directional dependence of the underground muon intensities according to the geometry of the surrounding rock and mountains (see \cref{fig:heatmap}). To cross-validate this dependence, we have performed two directional measurements in Site 0. The zenith and azimuth angles were selected by identifying the smallest slant depth and the related arrival direction, which corresponds to the highest underground muon intensity. For Site 0, this gives $\theta$ = \ang{34} and $\phi$ = \ang{270}. Due to the way the mountain profile is constructed, the MUTE azimuth angle follows a traditional cartesian plane. Therefore, \ang{270} corresponds to South (or \ang{180} in the compass quadrant system).

The first measurement was taken such that the detector was inclined by \ang{34} and rotated to point South. For the second measurement, the inclination angle remained the same, but the detector was rotated to face \ang{135} (or NW). This corresponds to a direction on the MUTE heatmap, which has a marked reduction in underground muon intensity compared to the first direction. Data were then taken and analyzed according to the previous sections, a count rate was computed for each direction, and the ratio of these was taken and converted to a percent reduction.

\begin{figure}
    \centering
    \includegraphics[width=8.5cm]{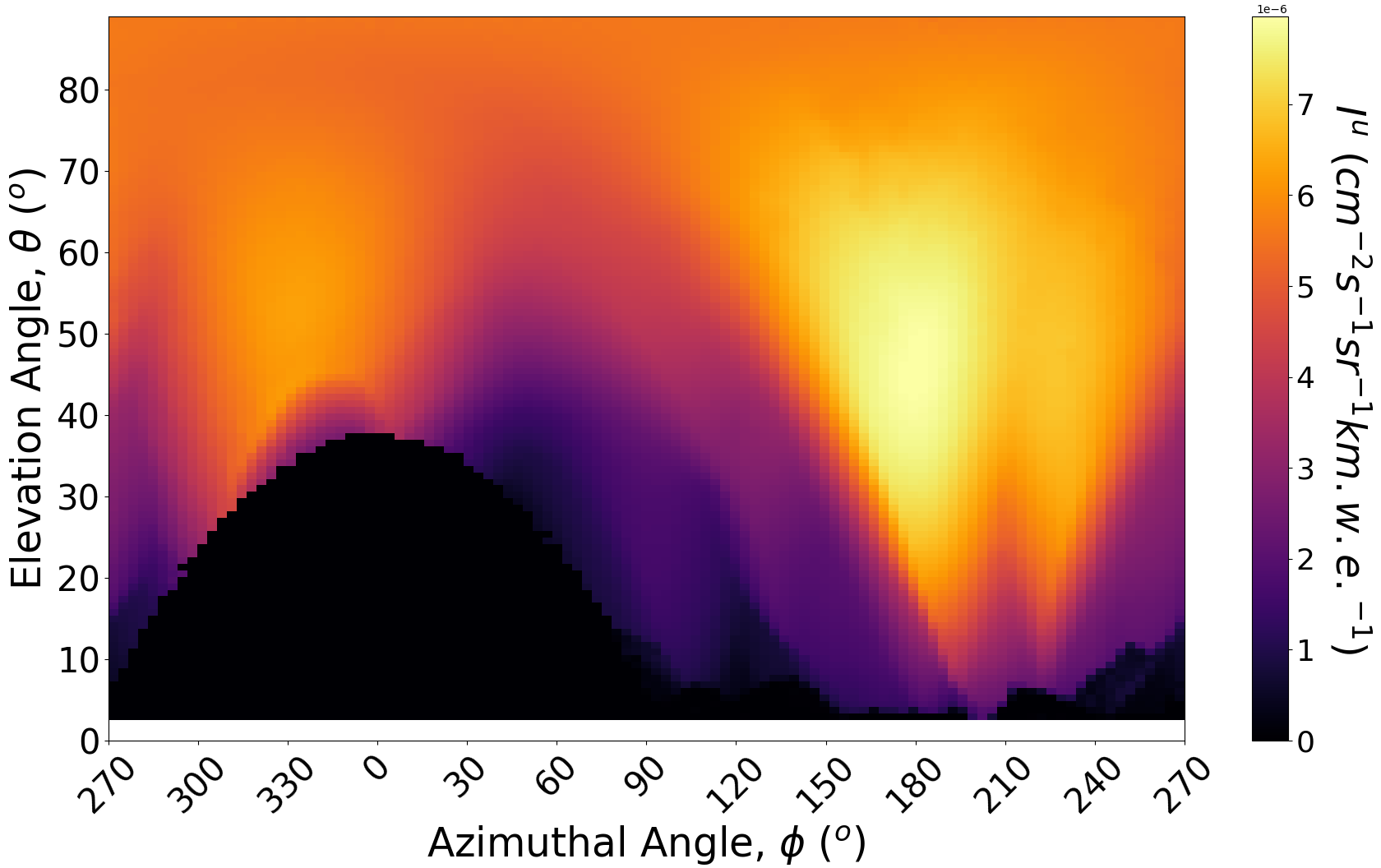}
    \caption{The detector acceptance folded into the MUTE heat map for the south-facing detector orientation. Here, \ang{0} indicates North, and the black patch indicates the arrival directions in which muons do not produce a coincidence trigger due to the detector geometry. Note that the y-axis is given as an elevation angle for a more straightforward visual interpretation.}
    \label{fig:DirectionalRatio}
\end{figure}

To directly compare these measurements to MUTE, Geant4 was used to simulate the inclined detector configuration and extract the detector response. Following the same weighting procedure detailed in \cref{sec:GeoCorrections}, a heatmap of the arrival-dependent muon intensities was reproduced for both directional measurements. \cref{fig:DirectionalRatio} shows the result of this process for the South facing configuration. The heatmap was then integrated over for both directions and the ratio was taken and converted to a percent reduction. These reductions in muon flux between the two directions for Site 0 yield 22.8 $\pm$ 2.1\,$\%$ and 24.0 $\pm$ 2.3\,$\%$ for measurement and simulation, respectively.


\section{Discussion and Conclusions}\label{sec:Discussion}
 
\begin{figure}[!htb]
    \centering
    \includegraphics[width=8.5cm]{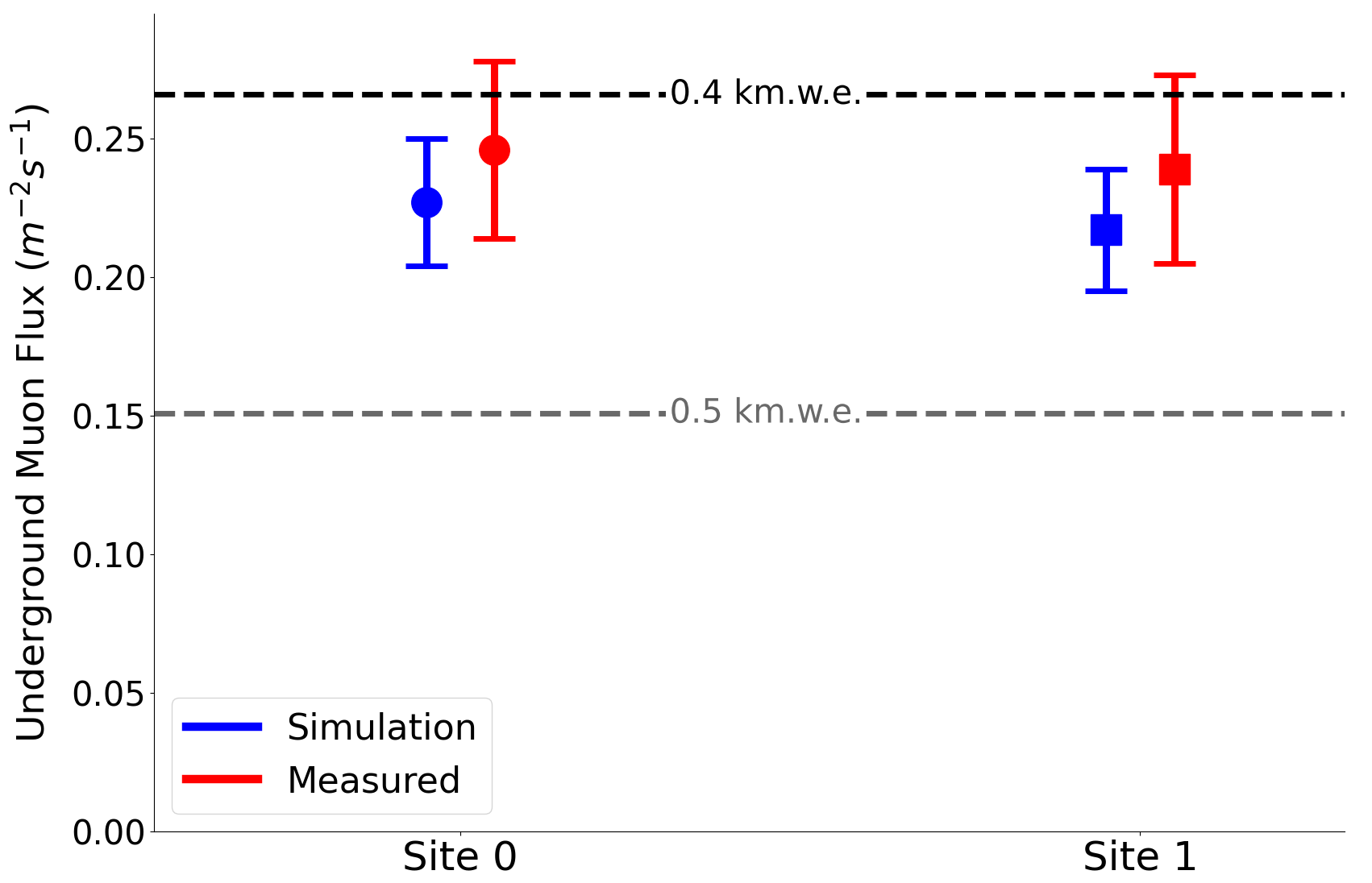}
    \caption{Comparison between the simulated and measured muon fluxes for Site 0 and Site 1. The circle markers correspond to Site 0, and the square markers correspond to Site 1. The error bars shown are both systematic and statistical for the measured flux and purely systematic for the simulated flux. The gray and black horizontal lines correspond to the flux equivalent of 0.5 and 0.4 km.w.e.\ flat overburdens, respectively.}
    \label{fig:sim_vs_exp}
\end{figure}

A summary of the results of this analysis can be seen in \cref{tab:resultsummary}. The muon flux measurements at the EEM demonstrate a close agreement with the simulation predictions, and a comparison can be seen in \cref{fig:sim_vs_exp}. The measured muon fluxes at Site 0 and Site 1 were 0.246 $\pm$ 0.020 $\pm$ 0.012 $\mu\text{/}$\unit[per-mode = symbol]{\square\metre\per\second} and 0.239 $\pm$ 0.025 $\pm$ 0.010 $\mu\text{/}$\unit[per-mode = symbol]{\square\metre\per\second}, respectively. These results are within 1$\sigma$ agreement of the simulated fluxes calculated using Daemonflux and MUTE. The agreements between measurement and simulation for both total flux and directional variations highlight the viability of these computational tools for predicting the underground muon flux at shallow facilities. The \SI{9}{\%} relative difference between measured and simulated total fluxes is consistent with past studies using these tools \cite{woodley2024cosmicraymuonslaboratories} and reflects the systematic uncertainties inherent in both the experimental setup and the overburden model. This result underscores the suitability of CURIE for hosting experiments that require moderate muon attenuation, offering a factor of $\approx700$ reduction in muon flux compared to sea level. This places CURIE in a favorable position relative to other shallow underground research facilities, particularly in the U.S., where options are limited.

The development of a new depth-intensity relationship allows for the direct comparison of its overburden with both flat and mountainous underground sites worldwide. This relationship can serve as a tool for researchers looking to evaluate potential underground locations for experiments. The derived equivalent depths for the EEM are consistent with those of other shallow underground labs, validating the accuracy of the simulation approach and the applicability of the model to similar facilities.

Moreover, the directional measurements conducted at Site 0 showed a strong dependence on the muon flux with respect to azimuthal and zenith angles, as predicted by MUTE simulations. The comparison between the directional fluxes facing South and Northwest yielded a \SI{23}{\%} reduction in muon flux, which was consistent with the simulation predictions and the importance of understanding the angular distribution of muons in designing and interpreting underground experiments.

In conclusion, the characterization of the EEM's muon flux and overburden depth confirms its potential as a valuable resource for low-background experiments. The facility provides a cost-effective and accessible option for experiments that do not require the extreme depth of larger underground labs. Future work will focus on performing extensive angular measurements and exploring muon spallation products inside the caverns, as well as their mitigation with additional experimental shielding. In particular, the muon-induced neutron background from the surrounding rock will be investigated using similar Monte Carlo methods, coupled with the results from MUTE and Daemonflux. Additionally, a more detailed understanding of the rock overburden is required to limit the geological systematics present throughout this work. A geochemical analysis would provide more precise data on the rock density and composition, leading to more accurate simulation results.

\section*{Acknowledgments}\label{sec:Acknowledgments}
This material is based upon work supported by a National Science Foundation Graduate Research Fellowship under Grant No. DGE-2137099, a U.S. Department of Energy Office of Science Grant No. DE-FG02-93ER40789, and the Colorado School of Mines via faculty start up funds and the ARCS Foundation. We thank Lee Fronapfel and Clint Dattel at the Edgar Experimental Mine for their support of our work. We would like to thank Dr. Dennis Soldin for their advice on muon simulation and fluxes. Finally, we would like to thank Dr. William Woodley for their invaluable input on the MUTE simulation work.


\bibliography{main.bib}


\end{document}